\documentclass{aastex}
\usepackage{epsf}
\usepackage{emulateapj5}

\newcommand{\HI}{H{\sc i}\ }

\shortauthors{de Blok et al.}
\shorttitle{LSB mass density profiles}
\begin{document}
\title{Mass Density Profiles of LSB Galaxies}
\author{W.J.G.~de~Blok\altaffilmark{1}}
\affil{Australia Telescope National Facility}
\affil{PO Box 76, Epping NSW 1710, Australia}
\email{edeblok@atnf.csiro.au}
\altaffiltext{1}{Bolton Fellow}
\author{Stacy S. McGaugh} 
\affil{Department of Astronomy, University of Maryland} 
\affil{College Park, MD 20742-2421, USA}
\email{ssm@astro.umd.edu}
\author{Albert Bosma} 
\affil{Observatoire de Marseille}
\affil{2 Place Le Verrier}
\affil{F-13248 Marseille Cedex 4, France}
\email{bosma@batis.cnrs-mrs.fr}
\and 
\author{Vera C. Rubin}
\affil{Department of Terrestrial Magnetism}
\affil{Carnegie Institution of Washington }
\affil{Washington, D. C. 20015, USA}
\email{rubin@dtm.ciw.edu}

\begin{abstract}

We derive the mass density profiles of dark matter halos that are
implied by high spatial resolution rotation curves of low surface
brightness galaxies.  We find that at small radii, the mass density
distribution is dominated by a nearly constant density core with a
core radius of a few kpc.  For $\rho(r) \sim r^{\alpha}$, the
distribution of inner slopes $\alpha$ is strongly peaked around
$\alpha = -0.2$.  This is significantly shallower than the cuspy
$\alpha \le -1$ halos found in CDM simulations.  While the observed
distribution of $\alpha$ does have a tail towards such extreme values,
the derived value of $\alpha$ is found to depend on the spatial
resolution of the rotation curves: $\alpha \approx -1$ is found only
for the least well resolved galaxies. Even for these galaxies, our
data are also consistent with constant density cores ($\alpha =0$) of
modest ($\sim 1$ kpc) core radius, which can give the illusion of
steep cusps when insufficiently resolved.  Consequently, there is no
clear evidence for a cuspy halo in any of the low surface brightness
galaxies observed.

\end{abstract}


\section{Introduction}

Low Surface Brightness (LSB) galaxies are dark-matter dominated
galaxies where the stellar populations only make a small contribution
to the observed rotation curves. It is therefore
straight-forward to compare the observed rotation curves of these
galaxies with those derived from numerical cosmological
simulations, where the dark matter is the dominant component. 

Early observation of dwarf and LSB galaxies showed that their rotation
curves rose less steeply than predicted by numerical simulations based
on the Cold Dark Matter (CDM) hypothesis
\citep{moore94,floresprimack94,edb_rot,mcg_nodm98}. In the CDM model, halos are
characterised by a steep central cuspy power-law mass density
distribution $\rho(r) \sim r^{\alpha}$. Initial simulations indicated
that $\alpha = -1$ \citep{NFW96}. More recent results indicate a more
extreme value $\alpha = -1.5$ \citep[e.g.][]{moore98,moore99,bul99}.
Rotation curves of dwarf and LSB galaxies, however, show a more
solid-body-like rise consistent with a mass distribution dominated by
a central constant-density core ($\alpha \simeq 0$), and hence are
inconsistent with the CDM predictions.  Similar conclusions have been
reached by \cite{salucci01} and \citet{salbor} for high surface
brightness disk galaxies.

The conclusions regarding LSB galaxies were based on \HI observations
with limited spatial resolution, and the data were in part affected by
beam smearing. Even though \citet{mcg_nodm98} showed that the steep
signatures of the rotation curves implied by CDM could not be hidden
by any reasonable amount of beam smearing, there were later
suggestions that the observed data \emph{were} consistent with the CDM
predictions
\citep{frankvdb,vdB&S} if proper beam smearing corrections were
applied.
\citet{SMT} published optical rotation curves of five LSB galaxies from the
\citet{edb_bmh96} sample and found that in several cases the
inner rotation curve slopes were somewhat steeper than found from the
\HI curves.  It is thus conceivable that the data could be reconciled
with CDM models once beam smearing corrections are properly taken into
account.  This is not borne out by improved data, as we show in this
Letter.

\section{The Data}

\citet{optcur_data} (MRdB), \citet{optcur_pap2} (dBMR) and
\citet{lsbopt_bosma} present high-resolution hybrid H$\alpha$/H{\sc i}
rotation curves of LSB galaxies, and show that a large fraction of
them are characterised by a slow, solid-body-like rise. They compare
(cuspy) CDM NFW halos \citep{NFW96} and pseudo-isothermal
(core-dominated) halos \citep[e.g.][]{beeg_phd} with the data and find
that the pseudo-isothermal model is statistically a much better fit to
the data.  Most NFW model fits suffer from systematic effects
resulting from the fitting program trying to reconcile $v(r)\sim r$
data with a steep $v(r) \sim r^{1/2}$ prediction.  The discrepancies
between \HI and optical data are found to be less severe than
initially suggested in \citet{SMT}: there is reasonable agreement
between new optical data and the old \HI curves in the majority of the
cases (MRdB).  The main conclusion drawn from the new data is that LSB
galaxies cannot be well fitted by CDM rotation curves.  Here we derive
the mass density profiles that are required to give rise to the
observed rotation curves. These can be compared directly to those
predicted by theoretical models.

We use the sample of  LSB and dwarf galaxy rotation curves described
in dBMR, MRdB and
\citet{lsbopt_bosma}. This sample includes the five LSB galaxies originally 
presented in \citet{SMT} that were re-derived in
dBMR. To this we have added the 4 LSB galaxies from the
Ursa Major sample of
\citet{marc_phd97} with $\mu_0(B) >22.0$ mag arcsec$^{-2}$ with
reliable \HI rotation curves.

\section{Results} 

In principle, one can invert an observed rotation curve to determine
the parent mass distribution.  In practice, this procedure can be
unstable for thin disks \citep{b&t} (but see \citealt{penny}).  For
LSB galaxies, the disk component is negligible since the potential is
dominated by the dark matter halo.  We therefore invert the observed
rotation curves assuming a spherical mass distribution, which is a
straightforward and robust procedure.  From
$\nabla^2 \Phi = 4\pi G \rho$
and 
$\Phi = -{GM}/{r}$
one can derive  the mass density $\rho(r)$:
\begin{equation}
4\pi G\rho(r) = 2 \frac{v}{r} \frac{\partial v}{\partial r} +
\left(\frac{v}{r}\right)^2,
\end{equation}
where $v$ is the rotation velocity and $r$ is the radius.  

Implicit in this procedures are assumptions: that these galaxies are all
dark matter dominated, that the gas motions are circular in a planar
disk and that the spectrum samples the nucleus and major axis.  We also
assume that the galaxies are symmetric.  (The latter is a good
assumption: MRdB, dBMR and \citet{lsbopt_bosma} omitted asymmetric galaxies
from the samples, whereas the effects of any mild, residual asymmetries
were incorporated in the errorbars.)

Here we ignore mass contributions of the stellar and gas
components. As long as these do not dominate the dynamics, as is the
case in LSB galaxies, this ``minimum disk'' assumption is a good
one. A minimum disk also produces an upper limit on the steepness of
the halo profile as inclusion of gas and stars will tend to flatten
the derived slopes.  See also dBMR and \citet{edb_rot}.  In a
forthcoming paper we show the results for non-minimum disk hypotheses,
which are not substantially different from the ones derived here.

\begin{figure} 
\begin{center}
\epsfxsize=0.9\hsize 
\epsfbox{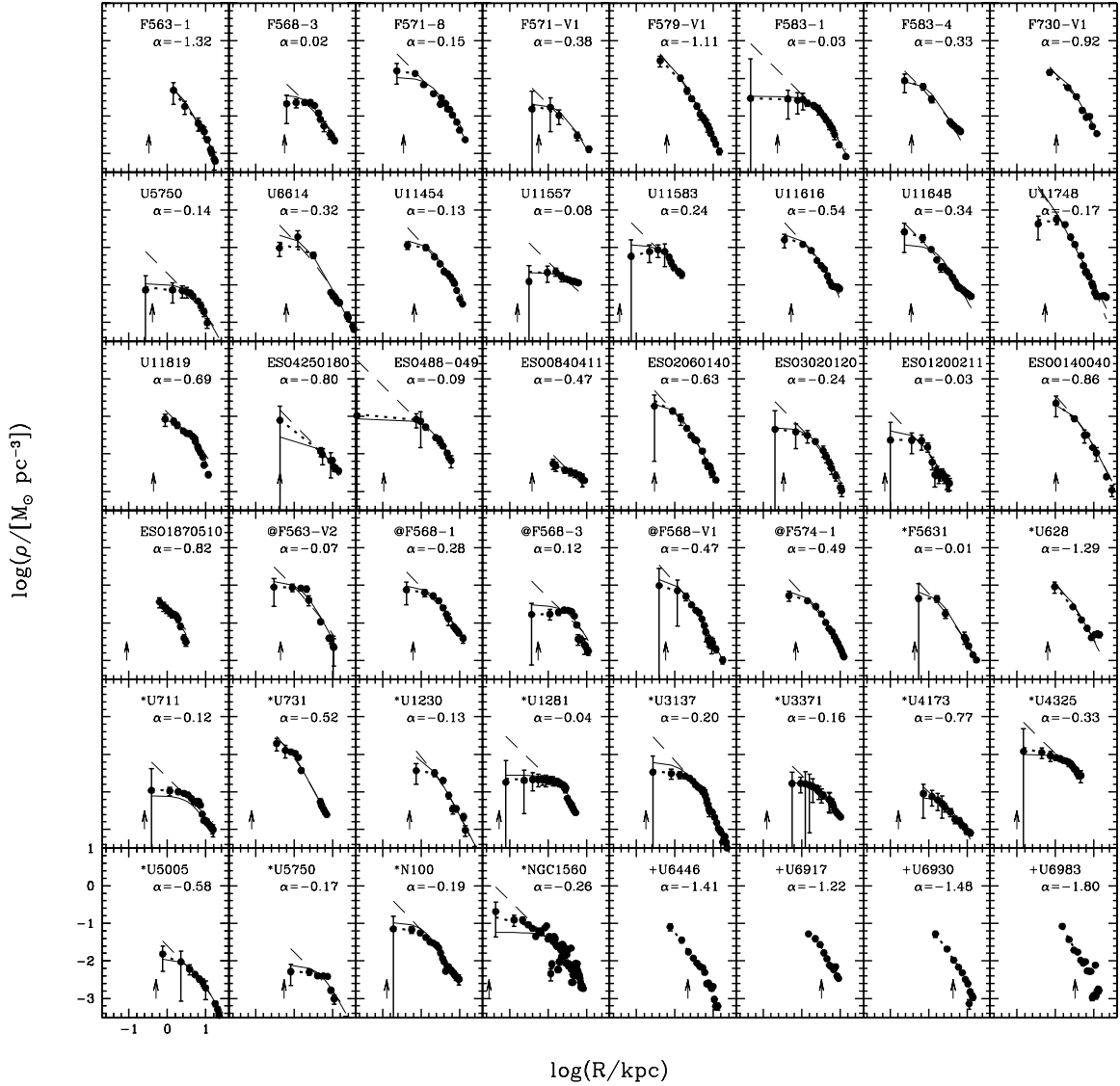}
\figcaption[deblok_fig1.ps]{Mass profiles of LSB galaxies (filled circles) 
derived from high-resolution rotation curves. The profiles can be
characterised by a steep $r^{-2}$ outer component, and a more shallow
inner component (``core''). Also shown are the mass density profiles
implied by the best-fitting minimum disk models from de Blok \& Bosma
(2001) and dBMR. Shown are the pseudo-isothermal halo (full line) and
the NFW halo (long-dashed line). We have also fit a power-law to the
inner shallow part (thick short-dashed line). The slope $\alpha$ is
given in the top-left corners of the panels. The arrows indicate the
size of the seeing disk. All panels are at the same scale (denoted in
bottom-left corner). Galaxies are labelled with their name, those from
de Blok \& Bosma (2001) are furthermore labelled with a star, those
from Verheijen (1997) with a plus-sign, and those derived from the
\citet{SMT} data with an ``@''-symbol.
\label{rho_min}}
\end{center} 
\end{figure}

Figure~\ref{rho_min} shows the derived mass density profiles.
Over-plotted are the profiles of the best fitting pseudo-isothermal
and NFW halo as listed in dBMR and
\citet{lsbopt_bosma}.  The errors on the data points are derived by
rigorously carrying through the errors in the rotation curve data
points.  The arrows indicate the size of the seeing disk for each
galaxy. (For the 4 Verheijen 1997 galaxies we indicate the radio 
beam size.)

The shape of the mass density profiles can generally be characterised
by two components: an outer one with an isothermal slope of $-2$ and a
more shallow one in the inner parts. After determining the
``break-radius'' where the slope changes most rapidly, we determine
the slope of the inner component using a weighted least-squares fit.
Table~\ref{tab:minslopes} collects the values of the inner slope.  The
range over which the power-law is fitted is indicated by the range
over which the fit is drawn in Fig.~\ref{rho_min}.  The uncertainty
$\Delta\alpha$ is determined by re-measuring the slope twice, once by
including the first data point outside the break-radius, and once by
omitting the data point at the break-radius. The maximum difference
between these two values and the original slope is adopted as the
uncertainty.  The isothermal mass model generally follows the derived
mass profile very well. The NFW models usually fail dramatically in
the inner parts. For the galaxies, the signature of the shallow inner
slope is usually already present well outside the seeing disk. Optical
``beam smearing'' can therefore not cause these shallow inner slopes.

Minimum disk does provide relevant limits on the inner slopes as is
particularly well illustrated by UGC 6614. This is a bulge-dominated
giant LSB galaxy, and we expect the bulge to contribute significantly
to the dynamics at small radius for any plausible stellar
mass-to-light ratio. However, even in the minimum disk case we find a
shallow slope $\alpha=-0.3$. This means that for any non-minimum disk
situation the dark matter must be depressed even further away from the
NFW case (i.e.\ the true slope must be even flatter).

\begin{figure} 
\begin{center}
\epsfxsize=0.5\hsize 
\epsfbox{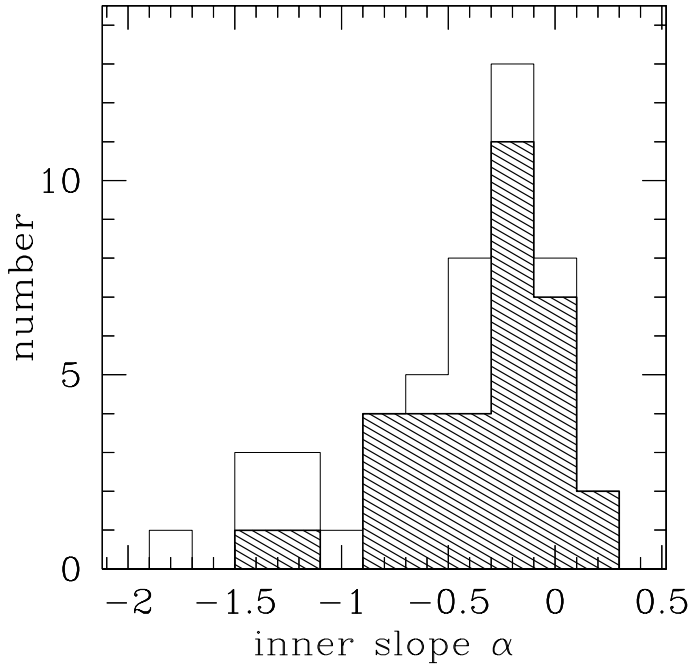}
\figcaption[deblok_fig2.ps]{Histogram of the values of the inner power-law slope
$\alpha$ of the mass density profiles presented in
Fig.~\ref{rho_min}. We distinguish between well-resolved (hatched
histogram) and unresolved (blank histogram) galaxies. The unresolved galaxies
generally have higher values of $\alpha$.
 \label{rho_min_hist}} 
\end{center} 
\end{figure}

Values for the inner slopes (Fig.~\ref{rho_min_hist}) are
asymmetric. In Fig.~\ref{rho_min_hist} we have distinguished between
galaxies where the profile is well-resolved and those where the
turn-over in the profile occurs at or within the seeing radius. The
most unresolved galaxies tend to have the most negative values of
$\alpha$. The resolved galaxies define a well-determined peak at
$\alpha = -0.2 \pm 0.2$ that is inconsistent with CDM predictions that
$\alpha = -1.5$.

One possible point of concern is the wing of steeper slopes extending
to $\alpha = -1.8$, where the extreme values originate from the
\citet{marc_phd97} UMa LSB profiles.  Does this mean that despite all
of the above, there are LSB galaxies that \emph{can} be well fitted
with CDM models? And does the tendency of the UMa LSB galaxies to have
steep slopes indicate systematic effects in the new LSB data? The
answer to both questions is negative, as the following analysis shows.

\begin{figure} 
\begin{center}
\epsfxsize=0.5\hsize 
\epsfbox{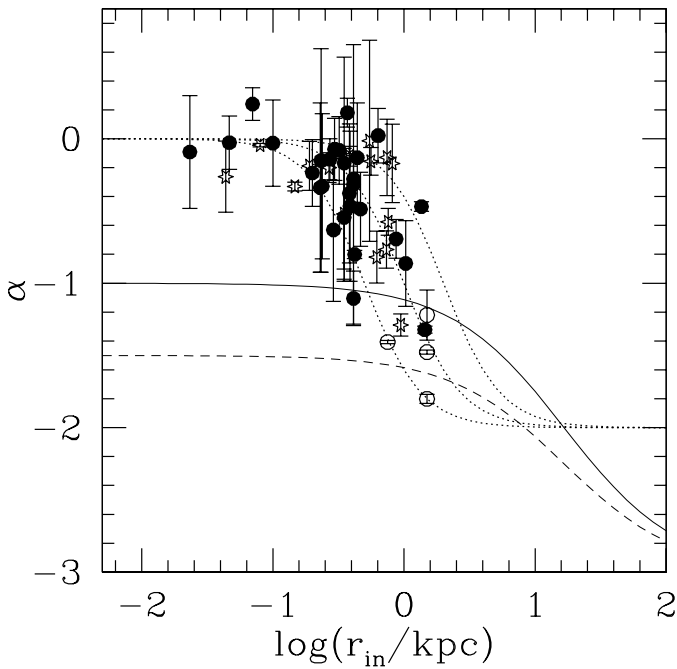}

\figcaption[deblok_fig3.ps]{Value of the inner slope $\alpha$ of the mass density
profiles plotted against the radius of the innermost point. Black dots
are from the dBMR sample, stars are from
the de Blok \& Bosma (2001) sample, open circles represent the four
LSB galaxies from the Verheijen (1997) sample. Over-plotted are the
theoretical slopes of a pseudo-isothermal halo model (dotted lines)
with core radii of 0.5 (left-most), 1 (canter) and 2 (right-most) kpc.
The full line represents a NFW model \citep{NFW96}, the dashed line a
CDM $r^{-1.5}$ model \citep{moore99}. Both of the latter models have
parameters $c=8$ and $V_{200} = 100$ km s$^{-1}$, which were chosen to
approximately fit the data points in the lower part of the diagram.
\label{sloperadius_min}} 
\end{center} 
\end{figure}

Table~\ref{tab:minslopes} lists the radius in kpc, $r_{\rm in}$, of the
innermost data point of each profile.  For the LSB sample
generally $r_{\rm in} < 1$ kpc.  For the UMa galaxies we find larger
$r_{\rm in}$: three of the four have $r_{\rm in}=1.5$ kpc.  In
Fig.~\ref{sloperadius_min}, we plot the values of $r_{\rm in}$ versus
the inner slope $\alpha$.  Also drawn are the logarithmic slopes as a
function of radius for pseudo-isothermal halos with core-radii $R_C =
0.5, 1, 2$ kpc, as well as a NFW model and a CDM $r^{-1.5}$ model (Moore
et al.\ 1999), both of the latter converging to a slope $\alpha=-3$ in
the (far) outer parts.  These two models are chosen to have parameters
$c=8$ and $V_{200} = 100$ km s$^{-1}$ to approximately match the four
UMa galaxies.  However, this choice is not critical. 

Galaxies with small values of $r_{\rm in}$ ($\lesssim 0.15$ kpc) show
clear evidence of a core ($\alpha \simeq 0$), whereas galaxies
with larger values of $r_{\rm in}$ exhibit steeper slopes.
Fig.~\ref{sloperadius_min} shows that distribution is consistent with
an isothermal halo with a core radius of a few kpc, whereas the NFW
and CDM models do not match the data at all.

Hence, only galaxies with small values of $r_{\rm in}$ measure the
core. Larger values sample a transition zone where the slope is
changing from $\alpha=0$ (center) to $\alpha=-2$ (outer isothermal
regions).  In the zone between $\sim 1$ and $\sim 10$ kpc the slopes
of the pseudo-isothermal and CDM models are approximately equal, so
large values of $r_{\rm in}$ might erroneously lead to the conclusion
that measured slopes are consistent with CDM.  The four UMa galaxies
(and some LSB galaxies) have $r_{\rm in}$ in this transition
zone and thus show steep slopes. We have modelled the beam smearing or
seeing effects potentially present in the optical data (discussed in a
forthcoming paper) and find that we can strongly exclude the
possibility that these affect the results down to resolutions of $\sim
0.1''$.  We thus predict that higher spatial resolution data (with
smaller values of $r_{\rm in}$) will also detect cores in the less
well-resolved galaxies.

Similar arguments apply to the beam smearing corrected \HI curves in
van den Bosch et al.\ (1999) and van den Bosch \& Swaters (2000).
With values $r_{\rm in} \sim 1$ kpc these data trace {\emph{not}} the
inner slope but instead the steep slope at the turnover of the
constant-density core.

\section{Conclusions}

Mass density profiles of LSB galaxies exhibit inner slopes that are
best described by a power-law $\rho(r) \sim r^{\alpha}$ with $\alpha =
-0.2 \pm 0.2$.  This result implies that halos of LSB galaxies are
dominated by cores. This result is inconsistent with the value $\alpha
= -1.5$ predicted by CDM models.  The steep slopes found for some LSB
galaxies arise when the innermost data point is sampling the
transition region between core and outer $\alpha=-2$ isothermal
region, not the core itself.  Our data are not consistent with CDM
predictions, but suggest that LSB galaxies have halos which contain
cores of radii of order 1 kpc.

\begin{deluxetable}{lrrrr}
\tabletypesize{\scriptsize}
\tablewidth{0pt}
\tablecolumns{5}
\tablecaption{Inner power-law slope $\alpha$\label{tab:minslopes}}
\tablehead{
\colhead{Galaxy}&\colhead{$\alpha$}&\colhead{$\Delta \alpha$}&\colhead{$r_{\rm in}$}&\colhead{$V_{\rm sys}^{\rm hel}$}\\
\colhead{}&\colhead{}&\colhead{}&\colhead{(kpc)}&\colhead{(km s$^{-1}$)}} 
\startdata
\sidehead{de Blok, McGaugh \& Rubin (2001)}
F563-1  & $-$1.32& 0.02  & 1.44& 3502  \\
F568-3  & $+$0.02& 0.19  & 0.64 & 5913  \\ 
F571-8  & $-$0.15& 0.77  & 0.23 & 3768  \\
F571-V1&  $-$0.38& 0.48  & 0.38 & 5721  \\
F579-V1 & $-$1.11& 0.19 &  0.41 & 6305 \\
F583-1  & $-$0.03& 0.19 &  0.05 & 2264  \\
F583-4  & $-$0.33& 0.50 &  0.24 & 3617  \\
F730-V1 & $-$0.92& 0.15 &  0.69 & 10714  \\
UGC 5750   & $-$0.14 &0.14  &  0.81& 4177  \\
UGC 6614   & $-$0.32& 0.97 &  0.42 & 6377  \\
UGC 11454  & $-$0.13& 0.38 &  0.44& 6628  \\
UGC 11557  &  $-$0.08& 0.23 & 0.32 & 1390  \\
UGC 11583 &   $+$0.24& 0.11 & 0.07 & 128  \\
UGC 11616 &  $-$0.54& 0.44 &  0.35 & 5244  \\
UGC 11648 &  $-$0.34& 0.59 &  0.23 & 3350  \\
UGC 11748 &  $-$0.17& 0.73 &  0.35 & 5265  \\
UGC 11819 &  $-$0.69& 0.13 &  0.87 & 4261  \\
ESO 0140040 & $-$0.86 &0.30 &  1.03& 16064  \\
ESO 0840411 & $-$0.47& 0.03 &  1.36   & 6200  \\  
ESO 1200211 & $-$0.03 &0.30 &  0.10 & 1314  \\
ESO 1870510 & $-$0.82 &0.18 &  0.62 & 1389  \\
ESO 2060140 & $-$0.63& 0.49 &  0.29 & 4704  \\
ESO 3020120 & $-$0.24 &0.23 &  0.20 & 5311  \\
ESO 4250180 & $-$0.80& 0.03 & 0.42 & 6637  \\
ESO 488-049 & $-$0.09& 0.39 &  0.02 & 1800  \\
\hline
\sidehead{de Blok \& Bosma (2001)}
F563-1 & $-$0.01 & 0.70 & 0.54 & 3502  \\
UGC 628   & $-$1.29 & 0.08 & 0.95& 5451  \\
UGC 711   & $-$0.12 &0.07  &  0.38& 1984 \\
UGC 731   & $-$0.52 &0.45  &  0.35& 637  \\
UGC 1230  & $-$0.13 &0.26   &  0.74& 3837  \\
UGC 1281  & $-$0.04 &0.01  &  0.08& 157  \\
UGC 3137  & $-$0.20 &0.10  &  0.27& 982  \\
UGC 3371  & $-$0.16 &0.10  &  0.56& 819  \\
UGC 4173  & $-$0.77 &0.13  &  0.73& 861  \\
UGC 4325  & $-$0.33 &0.03  &  0.15& 523  \\
UGC 5005  & $-$0.58 &0.09  &  0.76& 3830  \\
UGC 5750  & $-$0.17& 0.27 &  0.27 & 4177  \\
NGC 100  &  $-$0.19 &0.17 &   0.19& 841  \\
NGC 1560 & $-$0.26 &0.25  &   0.04& --36  \\
\hline
\sidehead{\citet{SMT}\tablenotemark{a}}
F563-V2& $-$0.07& 0.21  & 0.30 & 4312  \\   
F568-1& $-$0.28& 0.16  & 0.41 & 6524  \\
F568-3& $+$0.18& 0.10  & 0.38	&5913\\
F568-V1& $-$0.47& 0.52  & 0.39& 5768  \\
F574-1 & $-$0.49& 0.26  & 0.47 &  6889 \\
\hline
\sidehead{Verheijen (1997) UMa LSB galaxies}
UGC 6446 & $-$1.41 & 0.01  & 0.75&  644 \\
UGC 6917  & $-$1.22  &  0.17 & 1.50& 911  \\
UGC 6930 &  $-$1.48 &  0.01&  1.50&  777 \\
UGC 6983  & $-$1.80  &  0.03 & 1.50& 1082  \\
\enddata 
\tablenotetext{a}{These values are based on our re-derivation of 
the rotation curves using the method described in dBMR,
based on the raw data as published in \citet{SMT}.}
\end{deluxetable}

\end{document}